# Nickel-based phosphide superconductor with infinite-layer structure, BaNi$_2$P$_2$

Takashi Mine [1], Hiroshi Yanagi [1], Toshio Kamiya [1,2], Yoichi Kamihara [2], Masahiro Hirano [2,3], and Hideo Hosono [1,2,3,*]

[1] *Materials and Structures Laboratory, Tokyo Institute of Technology, 4259 Nagatsuta, Midori-ku, Yokohama 226-8503, Japan*

[2] *ERATO-SORST, JST, in Frontier Research Center, Tokyo Institute of Technology, 4259 Nagatsuta, Midori-ku, Yokohama 226-8503, Japan*

[3] *Frontier Research Center, Tokyo Institute of Technology, 4259 Nagatsuta, Midori-ku, Yokohama 226-8503, Japan*

**Abstract**

Analogous to cuprate high-$T_c$ superconductors, a NiP-based compound system has several crystals in which the Ni-P layers have different stacking structures. Herein, the properties of BaNi$_2$P$_2$ are reported. BaNi$_2$P$_2$ has an infinite–layer structure, and shows a superconducting transition at ~3 K. Moreover, it exhibits metallic conduction and Pauli paramagnetism in the temperature range of 4 – 300 K. Below 3 K, the resistivity sharply drops to zero, and the magnetic susceptibility becomes negative, while the volume fraction of the superconducting phase estimated from the diamagnetic susceptibility reaches ~100 vol.% at 1.9 K. These observations substantiate that BaNi$_2$P$_2$ is a bulk superconductor.

---

Footnotes:
(*)  Tel:    +81-45-924-5359
     FAX:   +81-45-924-5339
     E- mail:  hosono@msl.titech.ac.jp

## 1. Introduction

Cu- and other transition metal-based compounds are attractive for exploring high-transition-temperature (high-$T_c$) superconductors because it is considered that the high transition temperatures benefit from the strong electron correlation between $3d$ electrons in the transition metal atoms. Many cuprate high-$T_c$ superconductors have been discovered, including one reported in 1993, which has $T_c$ of 133 K [1]. On the other hand, typical $T_c$'s of other transition metal based compounds (including oxides and pnictides) are lower than those of cuprate high-$T_c$ superconductors. However, the discovery of new classes of superconductors, such as $Sr_2RuO_4$ [2], $Na_xCoO_2 \cdot yH_2O$ [3], electron-doped HfNCl [4], $Li_xNbO_2$ [5], has provided complementary insight to better understand the mechanism of superconductivity as well as clues for exploring higher-$T_c$ materials.

Recently, we have studied a series of quaternary compounds, which contain transition metal ions, La$T_M$O$Pn$ ($T_M$ = transition-metal cations such as Mn, Fe, Ni, and Co; $Pn$ = P and As), with the expectation that this series will be a new correlated electron system. In our studies, we have found new superconductors, LaFeO$Pn$ [6, 7] and LaNiO$Pn$ [8, 9], and an itinerant ferromagnet, LaCoO$Pn$ [10]. As shown in Fig. 1(a), La$T_M$O$Pn$ has a layered crystal structure where the positively charged La-O layers and negatively charged $T_M$-$Pn$ layers, which are composed of an edge-sharing network of $T_M Pn_4$ tetrahedra, are alternately stacked along the $c$ axis. It is thought that the $T_M$-$Pn$ layers sandwiched between the wider-gap La-O layers form magnetic and carrier conduction layers.

Ternary compounds containing transition metal ions, $AT_{M2}P_2$ ($A$ = Ca, Sr, Ba, and lanthanide cations), have the $ThCr_2Si_2$ structure and belong to the $I4/nmm$ space group. In addition, these ternary compounds show unusual physical properties such as intermediate valence states ($EuNi_2P_2$ [11]) and various magnetic properties, ranging from Pauli paramagnetism ($CaNi_2P_2$ [12]) to ferromagnetism ($LaCo_2P_2$ [11]), and to antiferromagnetism ($CaCo_2P_2$ [13]). Figure 1 (b) shows the crystal structure of $AT_{M2}P_2$, which has a layered structure similar to La$T_M$OP. The structure of the $T_M$-P layer is essentially the same as that in La$T_M$OP where the $T_M$-P layers are composed of an edge-sharing network of $T_M P_4$ tetrahedra. However, the wider-gap La-O layers in La$T_M$OP are replaced with the $A$ cation layers in $AT_{M2}P_2$. Thus, this crystal structure lacks a wider-gap insulating layer. Therefore, this structure may be regarded as an infinite-layer structure, which is analogous to cuprate high-$T_c$ superconductors. Because an infinite-layer cuprate, $(Sr_{1-x}Ca_x)_{1-y}CuO_2$, exhibits a high-$T_c$ of 110 K [14], we speculate that superconducting transitions are present in infinite-layer compounds composed of other transition metal cations, $ANi_2P_2$ and $AFe_2P_2$. Although to date, $AFe_2P_2$ ($A$ = Ca, Sr, Ba, La) and $ANi_2P_2$ ($A$ = Ca, La) have been examined, a superconducting transition has not been observed down to 1.8 K [12, 15]. Only $LaRu_2P_2$ with the $ThCr_2P_2$-type structure shows a superconducting transition at 4.1 K [15].

Here we report that a $ThCr_2Si_2$-type phosphide, $BaNi_2P_2$, exhibits a superconducting transition at ~3 K. Although Keimes et al. [16] have previously synthesized $BaNi_2P_2$ and reported its crystal structure, they did not report its electrical and magnetic properties. We synthesized ~90% pure $BaNi_2P_2$ samples, and measured their electrical and magnetic properties down to 1.9 K.

2. Experimental

Samples were prepared by a solid-state reaction of the starting materials, Ba (Johnson Matthey Company, 99.9%), P (Rare Metallic, 99.9999%), and Ni (Nilaco Corporation, 99.9%). A stoichiometric mixture of the Ba, P, and Ni powders was pressed into a pellet, and heated in an evacuated silica tube initially at 400 °C for 12 h and then at 1000 °C for 12 h. The sintered pellet was reground and subsequently pressed into a pellet, which was sintered at 1000 °C for 12 h. The resulting samples were characterized by high-power X-ray diffraction (XRD, D8 ADVANCE-TXS, Bruker AXS) with Cu Kα radiation, which detected trace amounts of impurity phases, $BaNi_9P_5$, $Ba(PO_3)_2$, and $BaNi_2(PO_4)_2$. Therefore, the crystal structure of $BaNi_2P_2$ and the compositions of the impurities were refined by the four-phase Rietveld method using the code TOPAS3 [17].

Electrical resistivity of the sintered pellets (apparent densities: ~63%) were measured in the temperature range from 1.9 to 300 K by the four-probe technique (using PPMS, Quantum Design). Sputtered Au films were used as ohmic contacts. Magnetic measurements were carried out using a vibrating sample magnetometer (VSM, using PPMS, Quantum Design). Temperature dependence of the magnetization was measured in a magnetic field at 10 Oe after zero-field cooling (ZFC) to the measurement temperatures.

3. Results and discussion

Figure 2 shows the powder XRD pattern of the purest sample obtained to date, which still shows diffraction peaks of $BaNi_2P_2$, $BaNi_9P_5$, $Ba(PO_3)_2$, and $BaNi_2(PO_4)_2$. Four-phase Rietveld analyses revealed that the obtained sample was mainly $BaNi_2P_2$, but contained ~9 vol.% of $BaNi_9P_5$, ~2 vol.% of $Ba(PO_3)_2$, and ~1 vol.% of $BaNi_2(PO_4)_2$. The obtained samples are dark gray and chemically stable in air.

Figure 3 shows the temperature dependence of the electrical resistivity ($\rho$) at an external magnetic field ($H$) of 0 Oe. The resistivity at 300 K was 2.8 mΩ·cm, and a metallic behavior was observed at temperatures down to ~3 K. The inset shows a magnified view in the temperature range of 1.9 – 10 K as a function of $H$. At $H$ = 0 Oe, $\rho$ dropped sharply at ~3 K, and the resistivity vanished at 2.7 K, implying a superconducting transition. The onset temperature where $\rho$ begins to drop decreases as $H$ increases, and the drop in $\rho$ vanishes at $H$ = 1000 Oe. These results suggest that the observed changes are due to a superconducting transition at ~3 K.

Figure 4 shows the temperature dependence of the mass magnetization ($M$) measured at 10

Oe after the ZFC. $M$ was as small as $\sim 1\times 10^{-4}$ emu/g, and was nearly independent of temperature at 4 – 300 K, implying Pauli paramagnetism in this temperature range. However, $M$ began to drop, became negative at ~3 K, and reached a large negative value of $-1.1 \times 10^{-1}$ emu/g at 1.9 K. These results, together with the zero resistance in Fig. 3, clearly indicate that the obtained sample exhibits superconductivity at temperatures below ~3 K. The field dependence of the magnetization (*M-H*) curve in the inset of Fig. 4 shows that the decrease in the negative magnetization is proportional to $H$ at $H$ < 150 Oe, and then increases to zero as $H$ increases up to 550 Oe. This behavior is similar to that observed in type-II superconductors. Furthermore, in this case the lower and upper critical magnetic fields were estimated to be $H_{c1}$ = ~150 Oe and $H_{c2}$ = ~550 Oe, respectively. The volume fraction of the superconducting phase estimated from the slope of the *M-H* curve at $H$ < 150 Oe was nearly 100%.

Next, the effects of the impurity phases were assessed. The impurity phase $BaNi_9P_5$ shows a temperature-independent Pauli paramagnetism [18], whereas $BaNi_2(PO_4)_2$ exhibits an antiferromagnetic transition at a Neel temperature of 24 K [19]. Rietveld analyses showed that the volume fractions of the impurity phases [~9 vol.% for $BaNi_9P_5$, ~1 vol.% for $BaNi_2(PO_4)_2$, and ~2 vol.% for $Ba(PO_3)_2$] were negligible compared to that of the superconducting phase. Consequently, we conclude that $BaNi_2P_2$ is a bulk superconductor below ~3 K. Although other Ni-based superconductors, *Ln*$Ni_2B_2C$ (*Ln* = Y, Tm, Er, Ho, and Lu) and $La_3Ni_2B_2N$, which are composed of tetrahedral Ni layers similar to LaNiO*Pn* and $BaNi_2P_2$, have been reported, each Ni ion is coordinated not by *Pn*, but by B ions [20, 21]. It is known in crystal chemistry and complex chemistry that $Cu^{2+}$ ions tend to take planar four coordinate structures, while Ni and Fe ions prefer to take tetrahedral structures, suggesting that a planar $Cu^{2+}$ structure is not a requisite for superconductivity, and such a transition metal-based tetrahedral layer is key to discovering new superconductors.

## 4. Summary

$BaNi_2P_2$, which belongs to the $AT_{M2}P_2$ system with the $ThCr_2Si_2$ structure, is regarded as an infinite-layer structure. $BaNi_2P_2$ shows a superconducting transition at ~3 K. Thus, $BaNi_2P_2$ is tentatively assigned as a type-II superconductor with a lower critical magnetic field of ~150 Oe and an upper field of ~550 Oe at 1.9 K. In the $AT_{M2}P_2$ system, the transition metal $T_M$ can be replaced with other transition metal ions, and the lattice parameters can also be controlled by replacing the *A* cation. These features provide a new platform to systematically survey the relationship among superconducting transitions, *d* electron configurations, and crystal structures. Hence, the discovery of higher-$T_c$ superconductors is anticipated in this and related material systems.

**References**

[1] A. Schilling, M. Cantoni, J. D. Guo, H. R. Ott, Nature 363 (1993) 56.

[2] Y. Maeno, H. Hashimoto, K. Yoshida, S. Nishizaki, T. Fujita, J. G. Bednorz, F. Lichtenberg, Nature 372 (1994) 532.

[3] K. Takada, H. Sakurai, E. Takayama-Muromachi, F. Izumi, R. A. Dilanian, T. Sasaki, Nature 422 (2003) 53.

[4] S. Yamanaka, K. Hotehama, H. Kawaji, Nature 392 (1998) 580.

[5] M. J. Geselbracht, T. J. Richardson, A. M. Stacy, Nature 345 (1990) 324.

[6] Y. Kamihara, H. Hiramatsu, M. Hirano, R. Kawamura, H. Yanagi, T. Kamiya, H. Hosono, J. Am. Chem. Soc. 128 (2006) 10012.

[7] Y. Kamihara, T. Watanabe, M. Hirano, H. Hosono, J. Am. Chem. Soc. 130 (2008) 3296.

[8] T. Watanabe, H. Yanagi, T. Kamiya, Y. Kamihara, H. Hiramatsu, M. Hirano, H. Hosono, Inorg. Chem. 46 (2007) 7719.

[9] T. Watanabe, H. Yanagi, Y. Kamihara, T. Kamiya, M. Hirano, H. Hosono, J. Solids State Chem. (2008), doi:10.1016/j.jssc.2008.04.033.

[10] H. Yanagi, R. Kawamura, T. Kamiya, Y. Kamihara, M. Hirano, T. Nakamura, H. Osawa, H. Hosono, (submitted).

[11] E. Mörsen, B. D. Mosel, W. Müller-Warmuth, M. Reehuis, W. Jeitschko, J. Phys. Chem. Solids 49 (1988) 785.

[12] W. Jeitschko, M. Reehuis, J. Phys. Chem. Solids 48 (1987) 667.

[13] M. Reehuis, W. Jeitschko, G. Kotzyba, B. Zimmer, X. Hu, J. Alloys Compd. 266 (1998) 54.

[14] M. Azuma, Z. Hiroi, M. Takano, Y. Bando, Y. Takeda, Nature 356 (1992) 775.

[15] W. Jeitschko, R. Glaum, L. Boonk, J. Solid State Chem. 69 (1987) 93.

[16] V. Keimes, D. Johrendt, A. Mewis, C. Huhnt, W. Schlabitz, Anorg. Allg. Chem. 623 (1997) 1699.

[17] Bruker AXS. TOPAS, version 3, Bruker AXS, Karlsruhe, Germany, 2005.

[18] J. V. Badding, A. M. Stacy, J. Solid State Chem. 87 (1990) 10.

[19] L. P. Regnault, J. Y. Henry, J. Rossat-Mignod, A. De Combarieu, J. Magn. Magn. Mat. 15-18 (1980) 1021.

[20] R. J. Cava, H. Takagi, H. W. Zandbergen, J. J. Krajewski, W. F. Peck Jr, T. Siegrist, B. Batlogg, R. B. Van Dover, R. J. Felder, K. Mizuhashi, J. O. Lee, H. Eisaki, S. Uchida, Nature 367 (1994) 252.

[21] R. J. Cava, H. W. Zandbergen, B. Batlogg, H. Eisaki, H. Takagi, J. J. Krajewski, W. F. Peck Jr, E. M. Gyorgy, S. Uchida, Nature 372 (1994) 245.

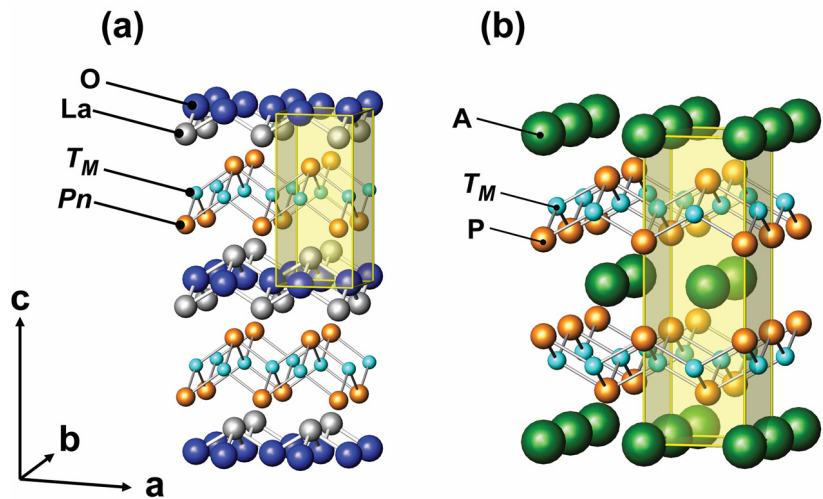

Fig. 1. (a) Crystal structure of La$T_M$O$Pn$. La-O layers and $T_M$-$Pn$ layers are stacked along the *c* axis. (b) Crystal structure of $AT_{M2}P_2$. Structure of the $T_M$-P layer is similar to that in La$T_M$OP.

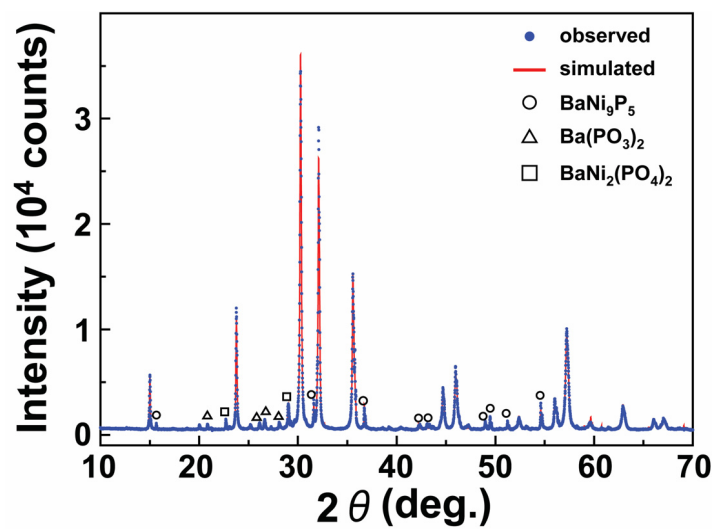

Fig. 2. XRD pattern of BaNi$_2$P$_2$ sample measured (blue circles) and simulated by the Rietveld method using the refined result (red line). Trace amounts of impurities, BaNi$_9$P$_5$, Ba(PO$_3$)$_2$, and BaNi$_2$(PO$_4$)$_2$, are detected, and their diffraction peaks are indicated by open circles, triangles, and squares, respectively.

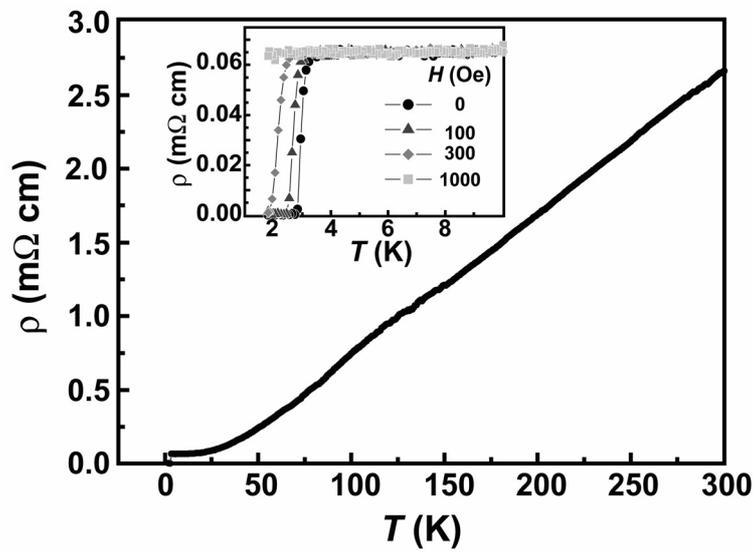

Fig. 3. Temperature (*T*) dependence of the electrical resistivity ($\rho$) at *H* = 0 Oe. Inset shows the $\rho$-*T* curves as a function of *H* magnified in the temperature range of 1.9-10 K.

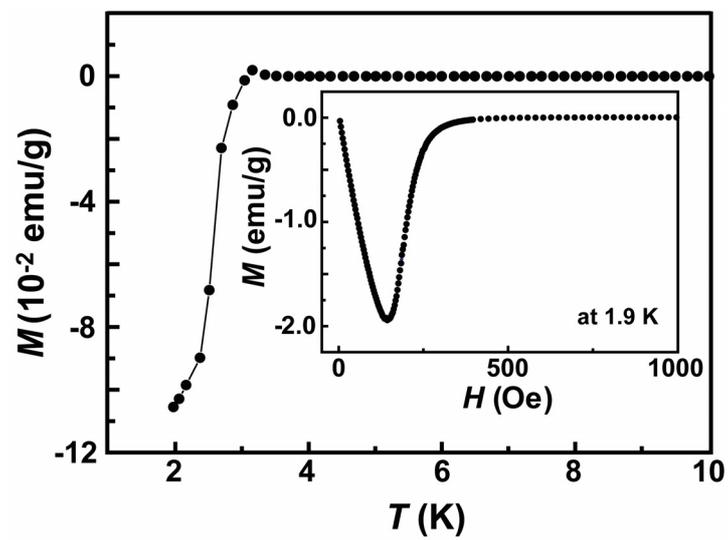

Fig. 4. Temperature (*T*) dependence of the mass magnetization (*M*) measured at 10 Oe after cooling to 1.9 K under a zero magnetic field. Inset shows the filed (*H*) dependence of *M* at 1.9 K.